\documentclass[12pt,epsf,amssymb]{article}
\usepackage{graphicx}
\usepackage{amsmath}
\usepackage{amsfonts}
\begin{document}

\def\a{\alpha}
\def\b{\beta}
\def\c{\varepsilon}
\def\d{\delta}
\def\e{\epsilon}
\def\f{\phi}
\def\g{\gamma}
\def\h{\theta}
\def\k{\kappa}
\def\l{\lambda}
\def\m{\mu}
\def\n{\nu}
\def\p{\psi}
\def\q{\partial}
\def\r{\rho}
\def\s{\sigma}
\def\t{\tau}
\def\u{\upsilon}
\def\v{\varphi}
\def\w{\omega}
\def\x{\xi}
\def\y{\eta}
\def\z{\zeta}
\def\D{\Delta}
\def\G{\Gamma}
\def\H{\Theta}
\def\L{\Lambda}
\def\F{\Phi}
\def\P{\Psi}
\def\S{\Sigma}

\def\o{\over}
\def\beq{\begin{eqnarray}}
\def\eeq{\end{eqnarray}}
\newcommand{\gsim}{ \mathop{}_{\textstyle \sim}^{\textstyle >} }
\newcommand{\lsim}{ \mathop{}_{\textstyle \sim}^{\textstyle <} }
\newcommand{\vev}[1]{ \left\langle {#1} \right\rangle }
\newcommand{\bra}[1]{ \langle {#1} | }
\newcommand{\ket}[1]{ | {#1} \rangle }
\newcommand{\EV}{ {\rm eV} }
\newcommand{\KEV}{ {\rm keV} }
\newcommand{\MEV}{ {\rm MeV} }
\newcommand{\GEV}{ {\rm GeV} }
\newcommand{\TEV}{ {\rm TeV} }
\newcommand{\sslash}[1]{\mbox{\ooalign{\hfil/\hfil\crcr{${#1}$}}}}
\newcommand{\Dsl}{\mbox{\ooalign{\hfil/\hfil\crcr$D$}}}
\newcommand{\ssl}{\mbox{\ooalign{\hfil/\hfil\crcr$\sigma$}}}
\newcommand{\sbsl}{\mbox{\ooalign{\hfil/\hfil\crcr$\overline{\sigma}$}}}
\newcommand{\nequiv}{\mbox{\ooalign{\hfil/\hfil\crcr$\equiv$}}}
\newcommand{\nsupset}{\mbox{\ooalign{\hfil/\hfil\crcr$\supset$}}}
\newcommand{\nni}{\mbox{\ooalign{\hfil/\hfil\crcr$\ni$}}}
\def\Z{\mathbb{Z}}
\def\R{\mathbb{R}}
\def\C{\mathbb{C}}
\def\Im{\mathrm{Im}}
\def\Re{\mathrm{Re}}
\def\vol{\mathrm{vol}}
\def\diag{\mathop{\rm diag}\nolimits}
\def\Spin{\mathop{\rm Spin}}
\def\SO{\mathop{\rm SO}}
\def\O{\mathop{\rm O}}
\def\SU{\mathop{\rm SU}}
\def\U{\mathop{\rm U}}
\def\Sp{\mathop{\rm Sp}}
\def\SL{\mathop{\rm SL}}
\def\tr{\mathop{\rm tr}}

\def\IJMP{Int.~J.~Mod.~Phys. }
\def\MPL{Mod.~Phys.~Lett. }
\def\NP{Nucl.~Phys. }
\def\PL{Phys.~Lett. }
\def\PR{Phys.~Rev. }
\def\PRL{Phys.~Rev.~Lett. }
\def\PTP{Prog.~Theor.~Phys. }
\def\ZP{Z.~Phys. }

\def\simgt{\mathrel{\lower2.5pt\vbox{\lineskip=0pt\baselineskip=0pt
           \hbox{$>$}\hbox{$\sim$}}}}
\def\simlt{\mathrel{\lower2.5pt\vbox{\lineskip=0pt\baselineskip=0pt
           \hbox{$<$}\hbox{$\sim$}}}}


\baselineskip 0.6cm

\begin{titlepage}

\begin{flushright}
UT-KOMABA/07-5

UT-07-13
\end{flushright}

\vskip 1.35cm
\begin{center}
{\large \bf
    Gauge Mediation with $D$-term SUSY Breaking
}
\vskip 1.2cm

Y. Nakayama,${}^{1}$ Masato Taki,${}^{2}$ Taizan Watari${}^{2}$ 
and T. T. Yanagida${}^{2}$

\vskip 0.4cm
${}^{1}${\it Institute of Physics, University of Tokyo,\\
      Komaba, Meguro-ku, Tokyo 153-8902, Japan}

\vskip 0.2cm

${}^{2}${\it  Department of Physics, University of Tokyo,\\
     Hongo, Bunkyo-ku, Tokyo 113-0033, Japan}

\vskip 1.5cm

\abstract{
We construct a gauge-mediation model with a $D$-term supersymmetry
(SUSY) breaking. $R$-symmetry breaking necessary for generating 
the SUSY standard-model gaugino masses is given by gaugino
condensation of a strongly coupled gauge theory in the hidden sector.
The energy scale of the strong dynamics of the hidden sector gauge 
theory should be around the messenger mass scale $M$, or otherwise 
perturbative calculations would be reliable and would lead to
 negative soft mass squared for squarks and sleptons.
Thus, all the mass scales are controlled by a virtually single parameter, 
$\sqrt{D}/M$. This model covers a very wide range of gravitino mass, 
$m_{3/2} \simeq 1\, \EV \mbox{--} 100 \, \TEV$.
Possible embeddings of the model in string theory are also discussed.
}
\end{center}
\end{titlepage}

\setcounter{page}{2}

\section{Introduction}

Supersymmetry (SUSY) can be broken either in the $F$-term or 
in the $D$-term (or both), but realistic models of SUSY breaking 
have been discussed almost exclusively in the $F$-term SUSY breaking.   
Two major issues of the $D$-term SUSY breaking scenario are 
 how to obtain $R$-symmetry breaking and 
how to avoid tachyonic masses (negative mass squared) 
for sfermions. 
Since a non-vanishing Fayet--Iliopoulos $D$-term parameter does 
not break an $R$-symmetry, something extra is necessary in order 
to obtain $R$-symmetry breaking gaugino masses in the SUSY 
standard-model (SSM).\footnote{For models with gaugino masses that
preserve $\U(1)_R$ symmetry, see e.g. \cite{Fox:2002bu}.}
Furthermore, in the gravity mediation with a generic K\"{a}hler 
potential,
soft mass-squared parameters of scalar fields from 
supergravity scalar potential,   
\begin{equation}
V = e^{K/M_P^2} (|F|^2 - 3|W/M_P|^2) + D^2/2g^2 \ ,
\end{equation}
are negative when the expectation value of $|W|^2$ is chosen 
so that the cosmological constant vanishes. Tachyonic squarks and 
sleptons imply that the color and electromagnetic U(1) symmetries 
would be broken in the vacuum.

Gauge group often becomes a product group, 
$G = G_1 \times G_2 \times \cdots$, in string theory compactification 
with D-branes. Only a part of the product gauge group is identified 
with $\SU(5)_{\rm GUT}$ or with the standard-model gauge group 
$\SU(3)_C \times \SU(2)_L \times \U(1)_Y$, and all other factors 
may become the hidden sector. When a U(1) factor is contained 
in the gauge group, it may have a non-vanishing $D$-term expectation 
value. If one of the hidden sector gauge groups is strongly coupled, 
and gaugino condensation is formed, $R$-symmetry is broken. 
Therefore, the only problem is how 
to mediate such an $R$-symmetry breaking to the standard-model sector.

In this article, we construct a model of gauge mediation 
in the $D$-term SUSY breaking scenario,\footnote{See \cite{Antoniadis:2007ta} and references therein for another line of models with $D$-term gauge mediation.} where the $R$-symmetry 
breaking from gaugino condensation is mediated through 
a messenger sector. The tachyonic mass problem may also be 
addressed in the gauge mediation because the tachyonic contribution 
from supergravity is small and negligible.
We find the SUSY-breaking mass squared can be positive in our 
model of gauge mediation only when the hidden sector gauge 
group is strongly coupled around the energy scale of the 
masses of the messenger fields.

This phenomenological constraint fixes one of freedoms 
in choosing parameters of the model. 
The $D$-term SUSY breaking scenario has three important parameters, 
namely, SUSY-breaking scale $\sqrt{D}/g$, the messenger mass scale $M$ 
and the $R$-symmetry breaking scale $\Lambda$. However, $\Lambda$ should be close to $M$ to avoid the negative mass squared for squarks and sleptons, but $\sqrt{D}$ and $M$ are chosen freely. This situation is 
similar to gauge mediation in the $F$-term SUSY-breaking scenario \cite{Giudice:1998bp}, where
$\sqrt{F}$ and $M$ are free. 
Depending on the value of $\sqrt{D}/M$, gravitino mass ranges 
from about $1\, \EV$ to $100 \, \TEV$ in our mediation model 
in the $D$-term SUSY breaking scenario. The gravitino mass can be 
that light because the SUSY breaking in a $D$-term of a U(1) symmetry 
is directly mediated to the visible sector through messenger fields 
charged under the U(1) symmetry. 
Light gravitino mass is very attractive 
from a phenomenological point of view.

We also discuss possible embeddings of the gauge mediation model 
in string theory. 
Fractional D$3$-branes located at a conifold singularity of a
Calabi--Yau three-fold \cite{Klebanov:2000hb} constitute a hidden sector while 
the standard-model sector is assumed to be realized on D$7$-branes 
on a four-cycle of the three-fold. 
D$3$--D$7$ strings become the messenger sector. 
Blow-up of the two-cycle at the conifold singularity breaks 
SUSY \cite{Dymarsky:2005xt} with a non-zero $D$-term. 

This paper is organized as follows.
In section 2, we present a field theory model of $D$-term gauge mediation. 
This phenomenological model is embedded in string theory in section 3.
Conclusions and discussions are found in section 4. In appendix A, we summarize the details of the two-loop computation for the SUSY scalar mass squared in the $D$-term gauge mediation. In appendix B, more examples of string theory embeddings can be found.

\section{Gauge mediation with $D$-term SUSY breaking}
\label{sec:phen}

In the $D$-term SUSY-breaking scenario, we have a U(1) gauge
symmetry, whose $D$-term expectation value $\vev{D}= g^2 \xi$ 
breaks SUSY. 
This U(1) symmetry is referred to as $\U(1)_{D}$ symmetry.
The $D$-term expectation value itself, however, does not break $R$-symmetry,
while any $R$-symmetries have to be broken down to $\Z_2$ so that 
Majorana gaugino masses are allowed. Thus, we introduce an $\SU(N)$ super
Yang--Mills multiplet, which leads to its gaugino condensation 
at low-energy: 
\begin{equation}
 \frac{1}{32\pi^2} \vev{2 \tr {}_N (\lambda^\alpha\lambda_\alpha)} 
  = \Lambda_N^3 \ ,
\end{equation}
where $\Lambda_N$ is the dynamical scale of the $\SU(N)$ gauge group. 
Here and hereafter, $\tr{}_N$ denotes the trace over the fundamental 
representation of this $\SU(N)$ gauge group. The gaugino condensation 
breaks $R$-symmetry down to $R$-parity.
This is probably the minimum construction of the hidden sector of 
the $D$-term SUSY breaking scenario.

In order for the SUSY breaking in the hidden sector to be
mediated to the visible particle-physics sector, messenger chiral
multiplets $\psi$ and $\bar{\psi}$ are introduced. They are in the 
{\bf 5} and $\bar{\bf 5}$ representations, respectively, of 
the SU(5)$_{\rm GUT}$ symmetry containing the standard-model gauge group 
$\SU(3)_C \times \SU(2)_L \times \U(1)_Y$. If the messenger chiral
multiplets carry non-vanishing charges of the U(1)$_D$ symmetry,  
the mass matrix of the complex scalars is of the form 
\begin{equation}
 V \sim 
\left( \begin{array}{cc} \psi^\dagger & \bar{\psi}^T  \end{array}\right)
\left( \begin{array}{cc} M^2 + g^2 \xi & 0 \\ 0 & M^2 - g^2 \xi
       \end{array}\right)
\left( \begin{array}{c} \psi \\ \bar{\psi}^*
       \end{array}\right) \ ,
\end{equation}
in contrast to a familiar form in gauge-mediation models in 
the $F$-term SUSY breaking scenario, 
\begin{equation}
  V \sim 
\left( \begin{array}{cc} \psi^\dagger & \bar{\psi}^T  \end{array}\right)
\left( \begin{array}{cc} M^2 & F \\ F^* & M^2 
       \end{array}\right)
\left( \begin{array}{c} \psi \\ \bar{\psi}^*
       \end{array}\right) \ .
\end{equation}
Here, $M$ is the SUSY-invariant mass of the messenger multiplets, 
$W = M \psi \bar{\psi}$ (We choose $M$ to be real). It has to be large enough, $M^2 > | \vev{D}|$, 
or otherwise $\psi$ (or $\bar{\psi}$) would develop an expectation value, 
breaking the symmetry of the standard model.
In order for the $R$-symmetry breaking to be mediated as well, 
we assume that the messenger multiplets $\psi$ and 
$\bar{\psi}$ are in the $(N,+1,{\bf 5})$ and $(\bar{N},-1,\bar{\bf 5})$ 
representations of the $\SU(N) \times \U(1)_D \times \SU(5)_{\rm GUT}$ 
gauge group.

\begin{figure}
\begin{center}
\input{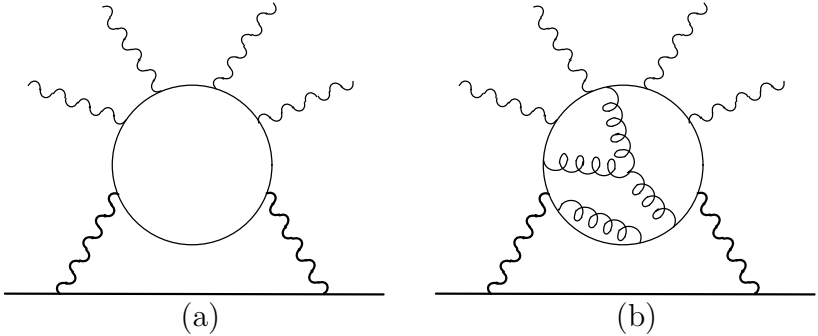}
\caption{Supergraphs contributing to squarks and slepton masses. 
Chiral multiplets in the visible sector are denoted by straight solid
 lines at the bottom, and the messenger chiral multiplets are the
 running in the loop. Gluon-like lines are used for the $\SU(N)$ 
vector multiplet, thick photon-like ones for the $\SU(5)_{\rm GUT}$ and 
ordinary photon-like ones for $\U(1)_D$. }
\label{fig:2loop}
\end{center}
\end{figure}

Soft SUSY-breaking mass squared is generated for scalar
particles in the visible sector after integrating out messenger fields. 
Two-loop diagrams in Figure~\ref{fig:2loop}(a) induce an effective operator 
in the K\"{a}hler potential,

\begin{equation}
 K_{\rm eff} = c \frac{g_{\rm {GUT}}^4}{16\pi^2}\frac{1}{16\pi^2}
\frac{|{\cal W}^\alpha {\cal W}_\alpha|^2}{M^6} f_if_i^{\dagger} \ ,
\label{eq:scalarmassKahler}
\end{equation}
where $c$ stands for coefficients of order unity and $g_{\rm GUT}$ 
for the gauge coupling constants of the SSM.
$f_i$ are matter chiral multiplets in the visible sector and 
${\cal W}^\alpha = \lambda^\alpha + \theta^\alpha D + \cdots$ 
the field strength superfield for the $\U(1)_D$ vector
multiplet.\footnote{We use a convention so that gauge kinetic terms are 
$\int d^2\theta \left[ \frac{1}{4g^2} {\cal W}^\alpha {\cal W}_\alpha 
+ \frac{1}{2g^2}\mathrm{tr}_N {\cal W}^{\alpha}{\cal W}_{\alpha} \right]
+ {\rm h.c.}$.}
Replacing the $\U(1)_D$ superfield ${\cal W}^\alpha$ by its
SUSY-breaking expectation value $\theta^\alpha \vev{D}$, 
we obtain soft masses squared of order 
\begin{equation}
 m_i^2 \sim \left[ \frac{\alpha_{\rm GUT}}{4\pi}\frac{\vev{D}^2}{M^3}
            \right]^2\ .
\label{eq:scalarmassOrder}
\end{equation}
Note that the soft mass squared is not generated at the order of 
$(\vev{D}/M)^2$, as opposed to the $F$-term SUSY breaking
scenario. Perturbative 2-loop calculation, whose details are found 
in appendix A, confirms that all the 
contributions of order $(\vev{D}/M)^2$ cancel out. 
The leading contribution of order (\ref{eq:scalarmassOrder}) turns out 
to be negative, as seen in appendix A. 
The first couple of terms in $|\vev{D}|/M^2$ expansion are all
negative, and one can further see by evaluating (\ref{smass}) numerically 
that all the squarks and sleptons have tachyonic masses as long as 
$M^2 >|\langle D \rangle| $.


The perturbative calculation in appendix A, however, is not reliable 
if the SU($N$) gauge theory in the hidden sector is strongly coupled 
around the threshold of the messenger fields. 
The SU($N$) gauge theory turns from a five-flavor SUSY ``QCD''
to a pure super Yang--Mills theory at the threshold, and its gauge 
coupling constant becomes strong immediately below the threshold 
for sufficiently large $N$.
Soft masses squared from all sorts of diagrams in Figure~\ref{fig:2loop}
(b) are of the same order as (\ref{eq:scalarmassOrder}); 
higher loop amplitudes are not suppressed because the loop factors 
$(g_N^2 N/16\pi^2)^n$ are of order unity, and effective operators 
with higher power of $|\tr {}_N ({\cal W}^\alpha {\cal W}_\alpha)|/M^3$ 
are just as important as the leading order operator 
(\ref{eq:scalarmassKahler}) because 
\begin{equation}
 \vev{2 \tr {}_N({\cal W}^\alpha {\cal W}_\alpha)} = 32\pi^2 \Lambda_N^3 \approx M^3
\label{eq:LambdaM}
\end{equation}
when the $\SU(N)$ interactions are strong just below the
threshold.\footnote{The dynamical scale $\Lambda$ of the $\SU(N)$ 
gauge theory cannot be chosen above the messenger mass scale for 
$N=2, 3$. This is because the $\SU(N)$ gauge theory is in the conformal 
window. For $N=4, 5$, $\Lambda$ may well, in principle, be much larger 
than the messenger mass scale $M$. A naive estimate of the 
leading contribution to soft mass-squared of squarks and sleptons 
is negative for $N=4$, but the leading calculable contribution of order 
$(\alpha_{\rm GUT}/4\pi)^2 D^4/M^6$ vanishes for $N=5$, and the sign of 
leading order non-vanishing contrbution, which is of order 
$(\alpha_{\rm GUT}/4\pi)^2 D^4/\Lambda^6$, is not calculable. 
Thus, it is not clear in the $M \ll \Lambda$ limit, 
whether the SUSY-breaking mass-sqared of squarks and
sleptons are negative or not. $N \geq 6$ is not compatible with 
perturbative unification and low-scale gauge mediation, and hence 
of less theoretical interest.
}
Since it is impossible to calculate all those contributions in 
such a strong-coupling regime, there is practically no way to know whether or not  
squarks and sleptons have tachyonic masses. 
Thus, we just assume that the sign is positive for all squarks and sleptons.\footnote{See \cite{IY,Ibe:2007wp} for a similar argument for the strongly coupled $F$-term gauge mediation.} 
Because of the naive-dimensional-analysis argument \cite{NDA}, 
eq.~(\ref{eq:scalarmassOrder}) still gives a reliable estimate 
of the soft SUSY-breaking masses squared even in this strong-coupling regime.

\begin{figure}
\begin{center}
\input{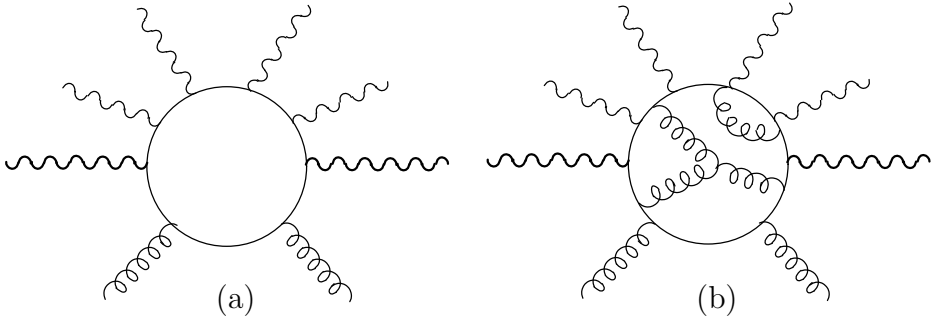}
\caption{Supergraphs contributing to gaugino masses.
See the caption of Figure \ref{fig:2loop} for the meaning of various 
lines.}
\label{fig:1loop}
\end{center}
\end{figure}

Gaugino masses of the standard-model gauge groups originate from 
one-loop diagrams of messengers in Figure~\ref{fig:1loop}(a). 
All diagrams in Figure~\ref{fig:1loop}(b), which are dressed by the SU($N$)
interactions also contribute to the gaugino masses by the same order 
of magnitude. 
Those diagrams induce an effective operator in the K\"{a}hler potential

\begin{equation}
  K_{\rm eff} = c' \, \frac{1}{16\pi^2} \frac{|{\cal W}^\alpha {\cal W}_\alpha|^2}{M^{10}}
                \tr {}_N ({\cal W}^{\alpha\dagger}{\cal W}^{\dagger}_{\alpha}) 
                \tr {}_{\rm GUT}({\cal W}^\alpha {\cal W}_\alpha) + {\rm h.c.} \ , 
\end{equation}
with a coefficient $c'$ of order unity. 
Field-strength superfields in $\tr {}_{\rm GUT}$ are those of the SSM. 
Thus, the gaugino masses of the SSM are of order 
\begin{equation}
  m_{\rm gaugino} \simeq 
 \frac{\alpha_{\rm GUT}}{4\pi} \, 
  \frac{32\pi^2 \Lambda^3_N}{M^3} \frac{\vev{D}^4}{M^7}, 
\end{equation}
where we have replaced $\tr {}_N ({\cal W}^\dagger {\cal W}^\dagger)$ 
by its expectation value.

Assuming that the $\SU(N)$ interactions are strongly coupled around the 
threshold, and hence (\ref{eq:LambdaM}), we see that the second factor 
$(32\pi^2 \Lambda_N^3/M^3)$ does not suppress gaugino masses relatively 
to sfermion masses.

Although the $D$-term SUSY breaking scenario at first seems to involve 
two independent parameters $\Lambda/M$ and $\sqrt{D}/M$, 
the former cannot be chosen freely for realistic models 
without tachyonic masses for squarks or sleptons.
Therefore, the spectrum predicted from the present model  
depends only on one parameter, $\sqrt{D}/M$. As we have already seen, 
SUSY-breaking parameters in the visible sector, 
\begin{equation}
 m_{\tilde{g},\tilde{w},\tilde{b}} \approx \frac{\alpha_{\rm GUT}}{4\pi}
                           \frac{\vev{D}^4}{M^7}\ , \qquad 
 m_{\tilde{q},\tilde{l}} \approx \frac{\alpha_{\rm GUT}}{4\pi}
                           \frac{\vev{D}^2}{M^3}\ ,
\end{equation}
depend on $\sqrt{D}/M$ in an interesting way.

The largest possible $\sqrt{D}/M$ corresponds to the lightest
gravitino mass possible. Messenger fields do not have tachyonic masses 
while $\sqrt{D}/M \simlt 1$, and at this limit, sfermion masses 
and gaugino masses are comparable 
(and both are supposed to be around the electroweak scale), and 
\begin{equation}
 M \sim \sqrt{D} \sim 100 \, \TEV\ , \qquad m_{3/2} \sim {\cal O}(\EV)\ .
\end{equation}
Theories with such a light gravitino is free from the constraint 
on the relic density of hot dark matter.\footnote{The dominant dark
matter may be axion or some other (possibly hidden) particle.}
Reheating temperature can be arbitrarily high, and in particular, 
thermal leptogenesis is not constrained in any ways by cosmological 
problems associated with gravitino.

If the SUSY breaking scale $\sqrt{D}$ is not as high as the 
messenger scale $M$, on the other hand, 
sfermions are much heavier than gauginos: 
\begin{equation}
 m_{\rm sfermion} \sim \left(\frac{M}{\sqrt{D}}\right)^4 \times m_{\rm gaugino} ,
\end{equation}
and the messenger scale and SUSY-breaking scale are
\begin{equation}
 M \sim \left( \frac{M}{\sqrt{D}}\right)^8 \times 100 \, \GEV\ , \quad 
 \sqrt{D} \sim \left( \frac{M}{\sqrt{D}}\right)^7 \times 100 \, \GEV \ ,
 \quad 
 m_{3/2} \sim \left(\frac{M}{\sqrt{D}}\right)^{14} \times {\cal O}(\EV)\ .
\end{equation}
for $m_{\rm gaugino} \simeq 100 \, \rm GeV$. The anomaly mediated contributions to gaugino masses are negligible
for $\sqrt{D}/M \simlt 1/10$ because the gravitino mass is no more 
than about $100 \, \TEV$. The spectrum of split SUSY \cite{splitSUSY} 
is realized without 
assuming a particular form of K\"{a}hler potential.

The size of the hidden-sector gauge group $N$ is arbitrary except that 
there should not be too much messenger fields. The messenger fields 
add $N$ pairs of chiral multiplets in the 
SU(5)$_{\rm GUT}$-${\bf 5}+\bar{\bf 5}$ representations. 
The gauge couplings of the standard-model gauge groups become 
asymptotic non-free, and may no longer be perturbative below the scale 
of gauge coupling unification. 
Since the gauge coupling unification is one of the most important 
motivations for low-energy SUSY, we do not want to lose it. 
If the messenger mass scale is of order 100 TeV, then the standard-model 
gauge couplings remain perturbative for $N \leq 6$. For the messenger 
mass scale of order $10^{12} \, \GEV$, $N \simlt 20$.

\section{$D$-term SUSY breaking models in string theory}
\label{ssec:simple}

It is rather straightforward to realize the idea of the $D$-term 
gauge mediation in string theory. Let us consider a Calabi--Yau orientifold
compactification of the Type IIB string theory. The standard-model gauge 
groups arise from a stack of five D$7$-branes wrapped on a holomorphic 
four-cycle $\Sigma$ in a Calabi--Yau manifold $X$. 
Quarks, leptons and Higgs super multiplets may arise from D$7$--D$7$
intersection. It is certainly a technically involved issue to find out 
an explicit geometry where exactly three generations of the matter
multiplets are obtained, but we just assume in this article that 
there is such a geometry because we focus on parts of geometry that 
is essential to a gauge mediation.

Suppose that the Calabi--Yau three-fold $X$ has a conifold singularity 
that is not contained in the four-cycle $\Sigma$ where the
standard-model gauge fields propagate.
When $N$ fractional D$3$-branes are at the singularity, 
a $\U(N)$ super Yang--Mills theory of ${\cal N} = 1$ SUSY 
is on the D$3$-branes.\footnote{Generalizations to the case with regular 
D3-branes are presented in appendix B.}
The $\SU(N)$ part can be identified with the hidden-sector $\SU(N)$, 
whose gaugino condensation breaks $R$-symmetry down to $R$-parity.

SUSY can be broken if $S^2$ at the conifold singularity 
is blown up. The SUSY-breaking vacuum energy is described in the 
effective field theory language as a non-vanishing Fayet--Iliopoulos 
parameter of the U(1) part vector multiplet. 
Thus, the U(1) part and its Fayet--Iliopoulos parameter 
can be identified with what we have needed in the $D$-term SUSY 
breaking model in section 2. 
In the string compactification with the finite internal volume, 
the Fayet--Iliopoulos parameter becomes a dynamical moduli. 
Once the K\"{a}hler moduli responsible for the Fayet--Iliopoulos
parameter is dynamical,
SUSY is restored by relaxing it down to $\xi = 0$.
We assume that the resolution K\"{a}hler moduli responsible for $\xi$ 
is fixed by some other mechanisms.\footnote{Local geometry has to be
made more complicated for this to happen. {\it c.f.} \cite{ADM}.} 

Open strings connecting those fractional D$3$-branes and the five D$7$-branes
wrapped on the four-cycle $\Sigma$ yield massive particles; their masses 
are proportional to the shortest distance between the blown-up $S^2$ 
and $\Sigma$. 
They are in the vector-like pair of representations, $(N,+1,{\bf 5})$ 
and $(\bar{N},-1,\bar{\bf 5})$, just as in section 2. Those particles 
are identified with the messenger sector chiral multiplets, $\psi$ and 
$\bar{\psi}$. 
By introducing fractional D$3$-branes at a conifold singularity, 
we have exactly what we need in the phenomenological model 
in section 2.
Since the low-energy effective theory of this Type IIB compactification 
is exactly the same as the field-theory model in section 2, all the 
results discussed in section 2 follow from this compactification.

%

\section{Discussions and Conclusion}

In this paper, we have proposed a new gauge mediation scenario based on the $D$-term SUSY breaking in the hidden sector. A crucial ingredient of our scenario is the gaugino condensation in the hidden sector, which yields the necessary $R$-symmetry breaking to produce the SSM gaugino mass. Notice that there is no $R$-axion problem \cite{Nelson:1993nf} because the $R$-symmetry is anomalous in our model.

Unfortunately, the perturbative two-loop computation shows that the SUSY scalar particles have negative masses-squared. Therefore, to apply the $D$-term gauge mediation successfully to our real world, the hidden sector should be strongly coupled near the thresholds of messengers. This requirement reduces the number of parameters of our model and makes our prediction even more concrete. 
Furthermore, this is rather a natural assumption because the hidden $\SU(N)$ 
gauge interactions become more asymptotic free after decoupling 
of the messenger fields. A possible explanation of the origin of the strong coupling at the threshold may be approximate conformal invariance above the messenger \cite{Ibe:2007wp}. 

Note that this $D$-term breaking scenario has virtually only one
gaugino-to-sfermion mass ratio and the gravitino mass.
For the largest $\sqrt{D}/M \sim 1$, sleptons are lighter than gauginos 
and the gravitino mass is of order eV. For a smaller value of
$\sqrt{D}/M$, sfermions are much heavier than gauginos, like in split 
SUSY, but the masses of those particles come from gauge
mediation. 

We have also discussed an embedding of $D$-term gauge mediation 
into the type IIB superstring theory. We have shown that the fractional 
D$3$-branes at the (resolved) conifold singularity together 
with the standard model flavor D$7$-branes realize the simplest 
$D$-term gauge mediation, given that the Fayet--Iliopoulos parameter is fixed. 
Generalized models presented in appendix B possess the gravity dual description. It would be interesting if we could find any evidence, from the supergravity viewpoint, for the positivity of the SUSY scalar 
mass-squared for squarks and sleptons, while strongly coupled nature 
of the $\SU(N)$ theory hinders the direct field theory computation.

\section*{Acknowledgements}
Y.N. thanks the Japan Society for the Promotion of Science for financial
support. 

\appendix
\section{Two-loop soft scalar mass squared from D-term SUSY breaking }
We report the two-loop soft scalar mass squared from gauge mediation with $D$-term SUSY breaking in this appendix. Under the $D$-term SUSY breaking, the mass squared for the messenger scalar takes the form as

\begin{eqnarray}
m^2 = \left(
  \begin{array}{cc}
   m_+^2 & 0  \\
   0 & m_-^2 
  \end{array} \right) = \left(
  \begin{array}{cc}
   m_f^2 + D & 0  \\
   0 & m_f^2 - D 
  \end{array} \right) \ ,
\end{eqnarray}
where  $m_f = M$ is the mass for the fermion.

Following the notation of  \cite{Martin:1996zb}, we find that the mass squared for the scalar SUSY particle is given by 
\begin{align}
\frac{2(16\pi^2)^2}{g^4}m^2_0 &= -2\langle m_+|m_+|0\rangle -2 \langle m_-|m_-|0\rangle -4m_+^2 \langle m_+|m_+|0,0\rangle - 4m_-^2\langle m_-| m_- | 0,0\rangle \cr
&- 4\langle m_f|m_f|0\rangle   + 8m_f^2 \langle m_f|m_f|0,0\rangle +4\langle m_+|m_f|0\rangle + 4\langle m_-|m_f|0\rangle \cr &+ 4(m_+^2-m_f^2)\langle m_+|m_f|0,0\rangle + 4(m_-^2-m_f^2)\langle m_-|m_f|0,0\rangle \ ,
\end{align}
where
\begin{align}
\langle m_1|m_2|m_3\rangle &= \int \frac{d^4q}{(2\pi)^4}\frac{d^4 k}{(2\pi)^4} \frac{1}{(q^2+m_1^2)(k^2+m_2^2)([k-q]^2 + m_3^2)} \cr 
\langle m_1|m_2|m_3,m_3\rangle &= \int \frac{d^4q}{(2\pi)^4}\frac{d^4 k}{(2\pi)^4} \frac{1}{(q^2+m_1^2)(k^2+m_2^2)([k-q]^2 +m_3^2)^2} \ 
\end{align}

After some algebra, we finally obtain
\begin{align}
\frac{2(16\pi^2)^2}{g^4} m^2_0 &= 
4\Biggl[\left(4M^2\right)\log\left(M^2\right)^2 
 + \log\left(M^2-D\right)\left(2\left(M^2-D\right) +\left(2M^2-D\right)\log\left(M^2-D\right)\right) \cr &
 + \log\left(M^2+D\right) \left(2\left(M^2+D\right) + \left(2M^2+D\right)\log\left(M^2+D\right)\right)  \cr &
 - 2\log\left(M^2\right)\left(2M^2 + \left(2M^2-D\right)\log\left(M^2-D\right) +\left(2M^2+D\right)\log\left(M^2 + D\right)\right) \cr &
 + 2D\mathrm{Li}_2\left(\frac{D}{D-M^2}\right) + \left(D+2M^2\right)\mathrm{Li}_2\left(\frac{D^2}{M^4}\right) \cr & - 2D\left(2\mathrm{Li}_2\left(\frac{D}{M^2}\right) + \mathrm{Li}_2\left(\frac{D}{D+M^2}\right)\right) \Biggr] \ ,  \label{smass}
\end{align}
where  $\mathrm{Li}_2(x) = \sum_{k=1}^\infty \frac{x^k}{k^2}$ is the dilogarithm function.
This expression is a special case of eqs.~(2.6--2.8) of \cite{PT}.
By expanding (\ref{smass}) with respect to  $D/M^2$, we obtain
\begin{align}
\frac{2(16\pi^2)^2}{g^4} m^2_0 = -\frac{14D^4}{9M^6} - \frac{76 D^6}{75M^{10}} - \frac{341 D^8}{490 M^{14}} - \cdots \ . \label{sm}
\end{align}
Note that the leading term is not  $O(D^2/M^2)$ but  $O(D^4/M^6)$ consistent with the effective operator analysis. The first three terms of the soft mass squared in (18) are all negative, and one can even see numerically that (17) is negative for any $M^2 > |\langle D \rangle| $.

\section{Generalized models in string theory }
\label{exam}

The hidden sector of the model in section \ref{ssec:simple} 
consists of $N$ fractional D$3$-branes on local (resolved) 
conifold singularity. 
This hidden sector can be extended to a system of 
$N+p$ fractional D$3$-branes and $p$ fractional D$3$-branes 
of the other type placed at the (resolved) conifold singularity. 
The gauge group of the hidden sector is 
$\SU(N+p) \times \SU(p) \times \U(1)_B \times \U(1)_{CM}$; 
both the $\U(1)_B$ and $\U(1)_{CM}$ symmetries are gauged 
because we consider a compact manifold $X$.\footnote{In fact, 
whether the vector fields of those $U(1)$ symmetries 
remain massless
depends on the details of the full compactification 
(see e.g. \cite{Buican:2006sn} for a recent discussion). 
Introduction of fluxes and the moduli stabilization make the problem 
even more complicated.}
The matter contents of the hidden and messenger sectors are 
summarized in Table~\ref{tab:contents}. 
This D-brane system is equivalent to $N$ D$5$-branes wrapped on 
the vanishing two-cycle at the conifold singularity and $p$ ordinary 
D$3$-branes.
The model in section \ref{ssec:simple} corresponds to the special case 
$p = 0$, and hence this class of models are generalization of the model 
in section \ref{ssec:simple}.

The U(1)$_B$ symmetry may have a non-vanishing Fayet--Iliopoulos
parameter $\xi_B$ \cite{Dymarsky:2005xt}. 
From the string theory perspective, it corresponds to blowing up 
a vanishing $S^2$ cycle at the conifold singularity. 
Once the moduli are fixed, 
then we know that this hidden-sector gauge theory breaks SUSY 
unless $p$ is an integral multiple of $N$ \cite{Dymarsky:2005xt}.

{\bf Vacuum Moduli in the Hidden Sector}

Matter fields in the hidden sector consists of 
four chiral multiplets, $A_\alpha$ ($\alpha=1,2$) and 
$B_{\dot{\alpha}}$ ($\dot{\alpha} = 1,2$) as in table \ref{tab:contents}. 
\begin{table}[tb]
\begin{center}
\begin{tabular}{c|cccccc}
        & $\SU(N+p)$& $\SU(p)$ & $\U(1)_B$ & $\U(1)_{CM}$ & $\SU(5)_{\rm
 GUT}$ & $[\SU(2)\times \SU(2)]_F$\\
 \hline
$A_{\alpha}$ & $N+p$ & $\bar{p}$ & $+2$ & $0$ & & $ 2\times 1$\\
$B_{\dot{\alpha}}$ & $\overline{N+p}$ & $p$ & $-2$ & $0$ & & $1 \times 2 $\\
\hline
$\psi$ & $N+p$ & & $+1$ & $+1$& ${\bf 5}$ & $ - $\\
$\tilde{\psi}$ & $\overline{N+p}$ & & $-1$ & $-1$ & $\bar{\bf 5}$ & $ - $\\
$\Psi$ & & $p$ & $-1$ & $+1$ & ${\bf 5}$ & $ -$\\
$\tilde{\Psi}$ & & $\bar{p}$ & $+1$ & $-1$ & $\bar{\bf 5}$ & $ - $\\
\end{tabular}
\end{center}
\caption{The matter contents of the SUSY breaking sector and messengers. Inclusion of messengers breaks the flavor $[\SU(2)\times \SU(2)]_F$ symmetry.}
\label{tab:contents}
\end{table}
They have a tree-level superpotential 
\begin{equation}
W_{\mathrm{tree}} = \frac{1}{\mu} (A_1B_1A_2 B_2 - A_1 B_2 A_2 B_1) \ .
\end{equation}

The hidden sector gauge theory (for $\xi_B = 0$) 
has a vacuum moduli space.
In a simplest case, $N=2$ and $p=1$, for example, 
the quantum vacuum moduli is parametrized by meson superfields
${\cal M}_{\alpha \dot{\alpha}} \sim A_\alpha B_{\dot{\alpha}}$ with 
a constraint
\begin{eqnarray}
{\cal M}_{11}{\cal M}_{22} - {\cal M}_{12}{\cal M}_{21} = \pm\sqrt{\Lambda_3^7\mu}  \ , 
\label{eq:dfmd-moduli}
\end{eqnarray}
which is the defining equation of (two copies of) the deformed conifold 
\cite{Klebanov:2000hb,Dymarsky:2005xt}. Here, the dynamical scale $\Lambda_3$ is defined by 
\begin{equation}
 \Lambda_3^7 = M^7 e^{-\frac{2\pi}{\alpha_3(M)}} \ , 
\end{equation} 
with the use of the gauge coupling constant $\alpha_{N+p}$ of $\SU(N+p)$ 
renormalized at the messenger mass scale $M$.
We notice that the $SU(3)$ gauge group shows a gaugino condensation 
\begin{equation}
 \frac{1}{32\pi^2} \vev{2 \tr{}_3 (\lambda^\alpha \lambda_\alpha)} = 
 \pm \sqrt{\frac{\Lambda_3^7}{\mu}}\ .
\label{eq:gg-cond}
\end{equation}

This moduli space corresponds to where the $p=1$ D$3$-brane is located 
in a deformed conifold defined by the equation \eqref{eq:dfmd-moduli}.
The non-vanishing Fayet--Iliopoulos parameter of the $\U(1)_B$ 
symmetry cannot be absorbed because all those meson fields 
${\cal M}_{\alpha \dot{\alpha}} \sim A_\alpha B_{\dot{\alpha}}$ 
are neutral under the baryonic U(1)$_B$ symmetry.
The vacuum energy, to leading order, is given by 
\begin{equation}
 V = \frac{g_B^2}{2} \xi_B^2 \ .
\end{equation}

{\bf Potential along the flat direction}

The leading order analysis shows a degeneracy of the non-supersymmetric
vacua along the flat directions \eqref{eq:dfmd-moduli}, 
but there is no reason to prevent the emergence of the potential 
along the radial direction of the
deformed conifold after the breaking of the SUSY. To see the
strongly coupled effects, we will study the gravity dual description. In
the gravity description, the model with the non-zero Fayet--Iliopoulos parameter is
represented by the so-called 
``warped-resolved-deformed conifold'' \cite{Butti:2004pk,Dymarsky:2005xt}. 

The pseudo-flat direction is then described by the probe D3-brane along the 
``warped-resolved-deformed conifold.'' 
The potential for such a probe brane comes from the nontrivial dilaton 
in the ``warped-resolved-deformed conifold'' 
\cite{Butti:2004pk,Dymarsky:2005xt}.

\begin{eqnarray}
V(t) = T_3 H^{-1}(t)(e^{-\phi(t)} -1) = 
  \frac{T_3}{\gamma}\frac{U^2}{e^{-\phi(t)}+1}\ 
\end{eqnarray}
in terms of the dilaton $\phi(t)$ and the warp factor $H(t)$,
where $t$ denotes the radial direction of the 
``warped-resolved-deformed conifold''. $U$ and $\gamma$ are related 
to the resolution/deformation parameter of the conifold respectively. 
Here we have also introduced the D3-brane tension $T_3$.  
The D3-brane is attracted toward $t=0$ and for large $t$, 
we have a very flat potential consistent with the field theory 
analysis.
 We note that the SUSY is broken even at $t=0$. 
For later purposes, let us study the potential value at $t=0$ and $t\to \infty$ more carefully. 
The dilaton here is normalized so that $\phi(\infty) = 0$ and the potential there is $V(t\to \infty) \sim \frac{1}{2} g_{B}^2 \xi_B^2$. 
On the other hand, for large $U$, $e^{-\phi(0)} \sim |U|^{3/4}$ 
\cite{Dymarsky:2005xt}. 
Since $U^2 = (\frac{\mu}{\Lambda_3})^{2/3} \frac{ \xi_B^2}{\Lambda_3^4}$ 
is typically large, the potential barrier is steep.

{\bf Inclusion of messengers}

The scalar potential along the flat direction discussed so far 
is modified as the probe $p=1$ D$3$-brane approaches the four-cycle 
$\Sigma$ where the $\SU(5)_{\rm GUT}$ gauge fields propagate.
Let the four-cycle be defined locally in the original conifold
by \cite{Ouyang:2003df}
\footnote{In the string compactification with the finite internal
volume, $M^2$ would be an open-string moduli. 
We again assume that the moduli $M^2$ is fixed by some other mechanisms.}
\begin{equation}
 {\cal M}_{11} = M^2 \ .
\end{equation}
Fields in the hidden sector and messengers have a superpotential
interaction \cite{Ouyang:2003df}
\begin{equation}
W = h \tilde{\psi} A_1 \Psi + g\psi B_1 \tilde{\Psi} + 
M_1 \psi \tilde{\psi} + M_2 \Psi\tilde{\Psi} \ .
\end{equation}
The parameters in the superpotential satisfy $M_1 M_2 = h g M^2$.
As the probe D$3$-brane approaches the four-cycle, 
${\cal M}_{11} \sim A_1 B_1 \sim M^2$, some fields in the messenger 
sector---D$3$--D$7$ open strings---become massless.
 When $\vev{A^{a=3}_{\alpha=1}} = \vev{B_{a=3,\dot{\alpha}=1}} \sim M$, 
the mass matrix in 
\begin{equation}
 W = \left(\tilde{\psi}_{a=3}, \tilde{\Psi}\right)
  \left(\begin{array}{cc}
   M_1 & h \vev{A^{a=3}_{\alpha=1}} \\ g \vev{B_{a=3,\dot{\alpha}=1}} & M_2
	\end{array}\right)
 \left(\begin{array}{c}
  \psi^{a=3} \\ \Psi
       \end{array}\right) + 
 M_1 (\tilde{\psi}^{a=1,2} \psi_{a=1,2}) 
\end{equation}
has a reduced rank.

Let the massless direction be 
\begin{equation}
 \psi_- = \cos \theta \; \psi^{a=3} + \sin \theta \; \Psi, \qquad 
 \tilde{\psi}_- = \cos \theta \; \tilde{\psi}^{a=3} +
 \sin \theta \; \tilde{\Psi} \ , 
\end{equation}
We use a common mixing angle $\theta$ for $\psi_-$ and $\tilde{\psi}_-$ 
because $h=g$ at all order in perturbation theory,\footnote{This is 
essentially due to the charge conjugation symmetry 
$\mathcal{M}_{12} \leftrightarrow  \mathcal{M}_{21}$ discussed in \cite{Klebanov:1998hh}. 
We have assumed that the inclusion of D$7$-branes does not break 
the symmetry.} and the $\SU(3)$ $D$-term condition ensures that 
$\vev{A^{a=3}_1} = \vev{B_{a=3, \dot{\alpha}=1}}$. 
The $D$-term potential of the $\U(1)_B$ and $\U(1)_{CM}$
symmetries are
\begin{equation}
 V_B +V_{CM} =  \frac{g_B^2}{2} \left( \xi_B  + \cos (2\theta)
\left(|\psi_-|^2 
    - | \tilde{\psi}_-|^2 \right) \right)^2 + 
\frac{g_{CM}^2}{2} \left(|\psi_-|^2 
    - | \tilde{\psi}_-|^2 \right)^2\ .
\end{equation}
The Fayet--Iliopoulos parameter $\xi_B$ induces negative mass squared
for either $\psi_-$ or $\tilde{\psi}_-$. 
$|\psi_-|^2 - |\tilde{\psi}_-|^2$ does not vanish  
at the minimum of the potential above, but the SUSY 
is not restored at the local minimum, with a remaining vacuum energy  
\begin{equation}
 V = \frac{g_B^2}{2} \xi_B^2 \frac{g_{CM}^2}{g_{CM}^2 + (\cos^2 2\theta)
  g_B^2} \ . 
\label{eq:U1U1}
\end{equation}
As long as $V(t=0)$ is less than this vacuum energy of the other 
local minimum, the vacuum we want does not have an instability. 
Since $V(t=0)$ is sufficiently smaller than the asymptotic value 
$V(t = \infty)$, there is a good chance that the $V(t=0)$ local minimum 
is more stable than this new local minimum.

{\bf Phenomenology}

We have seen so far that the generalized models in this appendix
break both SUSY and $R$-symmetry in the hidden sector, and 
those breaking are coupled to the messenger fields just like in 
the field-theory model of section \ref{sec:phen}. An important
difference, however, is that the energy scale of gaugino condensation 
is not the same as the dynamical scale in the models in this appendix.

Let us take a simple case $N=2$ and $p=1$ as an example. 
The $\SU(N+p)=\SU(3)$ gauge theory is in the conformal window above 
the messenger threshold scale \cite{CGT}, and the gauge coupling at the conformal fixed point is roughly $\alpha_3 \sim (3\pi/14)$. Below the messenger scale, 
the gauge theory with fewer matter fields quickly becomes strongly coupled. 
The dynamical scale is just below the messenger scale, $\Lambda_3 \sim M /4$.
We assumed that $32\pi^2 \Lambda^3/M^3 \approx {\cal O}(1)$ in section 
\ref{sec:phen} in order to avoid negative mass-squared for squarks and 
sleptons, and this assumption is completely valid. 

On the other hand, gaugino condensation in (\ref{eq:gg-cond}) and the meson masses involve 
an extra parameter $\mu$, whose typical value is at the string scale.
To assure the nonperturbative contributions to soft scalar mass squared, we take  $\mu \sim \Lambda$. The resulting phenomenology coincides with the one given in the main text.

\end{document}